\documentclass[5p,sort&compress]{elsarticle}
\usepackage{graphicx}
\usepackage{dcolumn,color}
\usepackage{bm}
\usepackage[percent]{overpic}
\usepackage{multirow,rotating}
\usepackage{epsfig}
\usepackage[english]{babel}

\newcommand{\er}{$\pm$}

\newcommand{\beq}{\begin{eqnarray}}
\newcommand{\eeq}{\end{eqnarray}}
\newcommand{\be}{\begin{eqnarray}}
\newcommand{\ee}{\end{eqnarray}}
\newcommand{\bea}{\begin{eqnarray}}
\newcommand{\eea}{\end{eqnarray}}

\newcommand{\bc}{\begin{center}}
\newcommand{\ec}{\end{center}}
\definecolor{lightgreen}{rgb}{0,0.4,0}
\definecolor{lightblue}{rgb}{0.5,0.5,1}
\definecolor{darkred}{rgb}{0.8,0,0}
\definecolor{darkgreen}{rgb}{0,0.4,0}
\definecolor{darkcyan}{cmyk}{1,0.3,0.3,0.3}
\definecolor{darkblue}{rgb}{0,0,0.6}
\definecolor{lightbrown}{rgb}{0.7,0.3,0.3}
\definecolor{darkbrown}{rgb}{0.5,0,0}

\newcommand{\rd}{\color{darkred}}
\newcommand{\bl}{\color{darkblue}}

\newcommand{\gr}{\color{darkgreen}}

\begin{document}

\begin{frontmatter}

\bibliographystyle{try}

\newcounter{univ_counter}
\setcounter{univ_counter} {0} \addtocounter{univ_counter} {1}
 \edef\HISKP{$^{\arabic{univ_counter}}$ }\addtocounter{univ_counter}{1}
 \edef\MURCIA{$^{\arabic{univ_counter}}$ } \addtocounter{univ_counter}{1}

\title{Scalar mesons and the fragmented glueball}

\author[label1]{Eberhard Klempt }

\address[label1]{Helmholtz--Institut f\"ur Strahlen-- und Kernphysik, Universit\"at Nu\ss allee 14-16, Bonn, Germany \vspace{-3mm}}
\date{\today}

\begin{abstract}
The center-of-gravity rule is tested for heavy and light-quark mesons.
In the heavy-meson sector, the rule is excellently satisfied. In the light-quark
sector, the rule suggests that the $a_0(980)$  could be
the spin-partner of  $a_2(1320)$, $a_1(1260)$, and $b_1(1235)$; $f_0(500)$ 
the spin-partner of $f_2(1270)$, $f_1(1285)$, and $h_1(1170)$; and 
$f_0(980)$  the spin-partner of $f_2'(1525)$, $f_1(1420)$, and $h_1(1415)$.  
From the decay 
and the production of light scalar mesons we find a consistent mixing angle 
$\theta^{\rm s}=(14\pm4)^\circ$. We
conclude that $f_0(980)$ is likely ``octet-like" in  SU(3) with a slightly larger $s\bar s$ content
and  $f_0(500)$ is SU(3) ``singlet-like" with a larger $n\bar n$ component. The 
$a_0(1450)$, $K^*_0(1430)$, $f_0(1500)$ and $f_0(1370)$ are suggested as nonet 
of radial excitations. The scalar glueball is discussed as part of the wave function 
of scalar isoscalar mesons and not as additional ``intruder". It seems not to cause supernumerosity.
 \end{abstract}

\end{frontmatter}

\section{Introduction}
Quantumchromodynamics (QCD) allows for the existence of a large variety of different states. SU(3)
symmetry~\cite{GellMann:1962xb} led to the interpretation of mesons and baryons as composed of constituent quarks~\cite{GellMann:1964nj},
as $q\bar q$ and $qqq$ states in which a colored quark and an antiquark with anticolor or
three colored quarks make up a color-neutral hadron~\cite{Fritzsch:1972jv}. Color-neutral objects can be formed 
as well as $(qq\bar q\bar q)$ tetra\-quarks \cite{Jaffe:1976ig}; the string between two quarks or a quark and an antiquark can be excited 
forming mesonic $(q\bar qg)$~\cite{Horn:1977rq} or baryonic $(qqqg)$~\cite{Barnes:1982fj}
hybrids. Glueballs with two or more constituent gluons 
may exist ($gg, ggg$)~\cite{Fritzsch:1972jv} as well as molecules generated by meson-meson~\cite{Meissner:1990kz}, 
meson-baryon~\cite{Bernard:1995dp} or baryon-baryon~\cite{Bogdanova:1971cr} interactions.    
These objects are all color-neutral, all may exist as ground states, all may have orbital excitations.  
A recent review of meson-meson and meson-baryon molecules can be found in~\cite{Guo:2017jvc} and of
further non-$q\bar q$ candidates in~\cite{Amsler:2020pdg}.  

This is a large variety of predicted states, and we may ask if all these possibilities are realized independently. 
In Ref.~\cite{Hanhart:2019isz} the question was raised, if - at least for heavy-quark mesons - quarkonia exist only 
below the first relevant $S$-wave threshold for a two-particle decay or if all quarkonia have at least a $Q\bar Q$ {\it seed}. 
In this letter we restrict ourselves to the discussion of scalar mesons in the light-quark sector. Here, all mesons fall above their threshold for $S$-wave 
two-particle decays. Indeed, the mostly accepted view is that we have
a nonet of mesonic molecules $a_0(980)$, $K^*_0(700), f_0(980)$, $f_0(500)$ -- also interpreted as te\-traquarks. 
But the two states $a_0(1450)$ and $K^*_0(1430)$ -- certainly above their threshold for $S$-wave 
two-particle decays -- are usually interpreted as $q\bar q$ states.
The two additionally expected scalar isoscalar $q\bar q$ states are supposed to mix with 
the scalar glueball thus forming the three observed mesons \,$f_0(1370)$, \,$f_0(1500)$, $f_0(1710)$~\cite{Amsler:1995tu,Amsler:1995td}.
Above $f_0(1710)$, no scalar isoscalar mesons are accepted as established in the Review of Particle Physics (RPP)~\cite{Zyla:2020zbs}. 

In this paper we study the possibility that all scalar mesons below 2.5\,GeV have a $q\bar q$ {\it seed}. They may acquire large
te\-traquark, molecular or glueball components~but~all scalar mesons can be placed into spin-multiplets containing 
tensor and axial-vector mesons with spin-parity $J^P=2^{++}$ or $1^{\pm}$. For the light scalar mesons
 we apply the center-of-gravity (c.o.g.) rule to the light tensor and axial vector mesons
to calculate the ``expected" mass of scalar mesons. Surprisingly we find, that the
predicted masses fall (slightly) below the mass of the light scalar mesons. 
We speculate that small $q\bar q$ components could be 
the seed of the light scalar mesons. Further, we investigate the possibility to determine the mixing
angle of the light scalar meson nonet. We argue that $a_0(1450)$, $K^*_0(1430)$, 
$f_0(1500)$, and $f_0(1370)$ can be accommodated as radial excitations. At the end, we discuss
the glueball. It seems not to invade
the spectrum as additional resonance but rather as component in the wave function of regular scalar mesons.

\section{The nonet of light scalar mesons}
The $\sigma$ meson is now firmly established as $f_0(500)$.  In the chiral limit, the pion is massless
and QCD is controlled by a single parameter $\alpha_s$ or $\Lambda_{\rm QCD}$. This is sufficient to generate a pole dynamically,
the $f_0(500)$ \cite{Hanhart:2008mx}. QCD dynamics, with only few free parameters, control the spontaneous breaking of chiral symmetry, 
confinement and the $f_0(500)$. There is no need for any $q\bar q$ or tetraquark component in its wave function. 
Inspite of this success, its nature as $q\bar q$ meson, tetraquarks or mesonic molecule is still a~topic~of~a controversial discussion. A survey of interpretations of the $f_0(500)$ can be found in \cite{Pelaez:2021dak}.
The existence of $\kappa$ or $K^*_0(700)$ is now certain as well \cite{Pelaez:2015qba,Pelaez:2020gnd,Pelaez:2020uiw}. 
Apparently the two resonances form, jointly
with $f_0(980)$ and $a_0(980)$, a nonet of meson resonances.  Here, we recall a few
different views.

Jaffe calculated the spectrum of light tetraquarks in the MIT bag mode~\cite{Jaffe:1976ig}. 
A nonet of light scalar mesons composed of two quarks and two antiquarks emerged 
while $q\bar q$ scalar mesons were found at masses well above 1\,GeV. 
This view was supported later by lattice calculations \cite{Alford:2000mm}.
Van Beveren, Rupp and collaborators~\cite{vanBeveren:1986ea} developed a 
unitarized non-relativistic meson model.
In the unitarization scheme for $S$-wave scattering, a $q\bar q$ resonance 
at 1300\,MeV with $J^{PC}=0^{++}$  quantum numbers originates from a confining potential. The resonance is coupled 
strongly to real and virtual meson-meson channels, creating a new pole in $\pi\pi$ scattering:
the  $f_0(500)$.  Similarly, $K^*_0(700)$, and $a_0(980)$ and $f_0(980)$
evolve from the unitarization scheme, no tetraquark configurations
are required. A similar model was suggested by Tornqvist and Roos \cite{Tornqvist:1995ay}.
In their model, the $f_0(980)$ and $a_0(980)$ are created by the $K\bar K\to s\bar s\to K\bar K$ 
interaction but owe their existence to $q\bar q$ states. The authors 
of Ref.~\cite{Santowsky:2020pwd} study in a Bethe-Salpeter approach the dynamical generation 
of resonances in isospin singlet channels with mixing between two and four-quark states. 
The authors conclude that the $f_0(500)$ wave function
is dominated by the $\pi\pi$ component; the tetraquark component is almost 
negligible, the quark-antiquark component contributes about 10\%. The first radial excitation
of the $f_0(500)$ is tentatively identified with the $f_0(1370)$. 
The two states $f_0(980)$ and $a_0(980)$ are very close in mass to the $K\bar K$
threshold. They are supposed to consist of two uncolored $q\bar q$ pairs and to be 
dynamically generated from $K\bar K$ interactions~\cite{Baru:2003qq}. Achasov {\it et al.}~\cite{Achasov:2020aun}
provide arguments in favor of the tetraquark picture of the light scalar mesons.

The two resonances $a_0(980)$ and $f_0(980)$ are both seen in lattice calculations
\cite{Dudek:2016cru,Briceno:2017qmb}. Their unusual pole structure is argued to
point at a strong $K\bar K$ molecular component. The $f_0(500)$ is ``stable" 
(the pion mass is 391\,MeV). Using the Weinberg criterium~\cite{Weinberg:1962hj} discussed below, the authors
find a molecular $\pi\pi$ component of about 70\%. 
The authors of ref.~\cite{Alexandrou:2017itd} performed lattice studies of the 
quark content of the $a_0(980)$ in two-meson scattering.  It is suggested that the
$a_0(980)$ is a superposition of $q\bar q$ and tetraquark.

Rupp, Beveren, and Scadron reject the hypothesis that 
a strict distinction can be
made between ``intrinsic'' and ``dynamically generated'' states  \cite{Rupp:2001zx},
and Jaffe argued against the possibility to define a ``clear distinction 
between a meson-meson molecule and a $qq\bar q\bar q$ state'' \cite{Jaffe:2007id}. 
Such a distinction is certainly possible only in the proximity of an $S$-wave threshold~\cite{Guo:2017jvc}.

\section{The center-of-gravity rule}
In the absence of tensor and spin-spin forces, the mass of the singlet heavy
quarkonium states is given by the weighted average of the triplet states~\cite{Lebed:2017yme}:\vspace{-2mm}

\begin{eqnarray}
\hspace{-8mm}M_{h_{Q}(nP)} =\frac19 \left(5 M_{\chi_{Q2}(nP)}\right.
%& \\ &\hspace{-12mm} 
\left.+ 3 M_{\chi_{Q1}(nP)} + M_{\chi_{Q0}(nP)}\right) 
\end{eqnarray}

The differences between the measured and the predicted masses of the $h$ masses
are  shown in Table~\ref{cog}. The masses were taken from the Review of Particle Physics 
(RPP)~\cite{Zyla:2020zbs}. In the sector of heavy quarkonia, the c.o.g. rule is excellently satisfied.

\begin{table}[pt]
\vspace{-2mm}
\caption{\label{cog}The center-of-gravity rule for heavy quarkonia 
\vspace{-2mm}}
\begin{center}
\renewcommand{\arraystretch}{1.3}
\begin{tabular}{cccc}\hline\hline
               &\phantom{zzz}& \multicolumn{2}{c}{\rm measured - predicted}\\\hline
$\delta\,M_{h_{c}(1P)}$:  &&    $(0.08\pm 0.61)$&MeV    \\    
$\delta\,M_{h_{b}(1P)}$:  &&     \hspace{-3mm}$-(0.57\pm 1.08)$&MeV      \\
$\delta\,M_{h_{b}(2P)}$:  &&  \hspace{-3mm}$-(0.4\pm 1.3)$&MeV \\ \hline\hline
\end{tabular}
\end{center}
\vspace{-2mm}
\end{table}

We next test the rule for charmed mesons. However, the mixing angle for the two 
charmed mesons $D_{s1}(2460)$ and $D_{s1}(2536)$ is not known. In the heavy-quark limit, 
we expect no mixing. We test the c.o.g. rule
with this assumption:
\begin{eqnarray}
M_{D_{s1}(2536)} =&\hspace{-15mm}\frac19 \left(5 M_{D^* _{s2}(2573)} +\right.\nonumber\\
& \left. 3 M_{D_{s1}(2460)} + M_{D^* _{s0}(2317)}\right)
\end{eqnarray}
The difference between the left-hand $(2535.11\pm 0.06)$\,MeV
and right-hand $(2504.6\pm 0.7)$\,MeV side is 30\,MeV. 
A finite mixing angle would be needed to satisfy the c.o.g. rule.
We note that  according to the Weinberg criterion discussed below, $D^* _{s0}(2317)$ and $D_{s1}(2536)$
have a large molecular component and could even be purely molecular states \cite{Matuschek:2020gqe}.
Here we assume that they have at least a $c\bar s$ seed.
Too little is known for the $D$-meson excited states to test the c.o.g. rule.  

We now apply the c.o.g. rule for light mesons to predict the mass of scalar
mesons. We are aware of the fact that this is dangerous and possibly misleading,
in particular for the calculation of the $f_0(500)$ and $K^*_0(700)$ states. 
These are very broad states, and the assumption that they have similar
internal dynamics as the tensor and axial vector mesons must be wrong. 
Certainly, we do not know what the impact is of meson-meson loops. Nevertheless, it seems
legitimate to us to explore the borders of applicability of the c.o.g. rule. 
%Naively, we may expect deviations to be enhanced by the factor $M_Q/M_q=1370/3.5$, hence $\approx 200$\,MeV
%instead of the $\approx 0.5$\,MeV in charmonium.
 
We start with the equation
\beq 
M_{f_0} = 9 M_{h_1} - 5 M_{f_2}  - 3 M_{f_1},
\eeq  
and use $M_{h_1}$=(1170\er20)\,MeV, $M_{f_2}$=(1275.5\er 0.8), and $M_{f_1}$ =(1281.9\er0.5)\,MeV.
We find a scalar mass of $M_{f_0}$=(307 \er180)\,MeV where the uncertainty is given by
the uncertainty the $h_1(1170)$ mass. In the other cases, the errors are added quadratically. 
For mesons with hidden strangeness, we have the masses 
$M_{h_1'}$=($1416\pm8$)\,MeV, $M_{f_2'}$=(1517.4\er 2.5), and $M_{f_1'}$ =(1426.3\er0.9)\,MeV, and
we find a scalar mass of $M_{f_0}$= (878\er73)\,MeV. 
Likewise, we estimate the expected mass of the scalar isovector meson from 
\begin{equation}
M_{a_0}= 9 M_{b_1} - 5 M_{a_2} - 3 M_{a_1}.
\end{equation}
We use the masses $M_{b_1}$=(1229.5\er3.2)\,MeV, $M_{a_2}$= (1316.9 \er0.9)\,MeV, 
and  $M_{a_1}$=(1230\er40)\,MeV. Thus we obtain a scalar mass of 
$M_{a_0 }$=(791\er 124)\,MeV. 

In the strange-quark sector, we have the well known $K^*_2(1430)$ at (1427.3\er1.5)\,MeV 
and two axial vector mesons, $K_1(1280)$ and $K_1(1400)$ with masses  of (1253\er7)
and (1403 \er7)\,MeV, respectively. The scalar meson mass can be estimated to
\begin{equation}
M_{K^*_0}= 9 M_{K_{1B}} - 5 M_{K^*_2} - 3 M_{K_{1A}}.
\end{equation}
The two observed resonances $K_1(1270)$ and $K_1(1400)$ are mixtures of $K_{1A}$ and $K_{1B}$.
A recent analysis of the mixing parameters suggests a mixing angle of 
$-(33.6\pm4.3)^\circ$~\cite{Divotgey:2013jba} and masses of $M_{K_{1A}}$=(1360\er5)\,MeV and
$M_{K_{1B}}$= (1310\er5) MeV. With these values we derive a scalar mass 
$M_{K^* _0}$= (574\er 40)\,MeV.

\begin{table}[pt]
\vspace{-2mm}\caption{\label{assign}Assignment of light scalar mesons to spin-multiplets. All masses are in MeV.
The first line gives the predictions based on the center-of-gravity rule of light tensor,
axial-vector and scalar mesons. The second line gives the name, the third one the best mass values
of our assignment to be spin-partners of $a_2(1320)$, $f_2(1270)$, $f_2(1525)$, and $K^*_2(1430)$.  
The assignments suggested
by the Particle Data Group is given in the last line.\vspace{2mm}
}
\renewcommand{\arraystretch}{1.3}
\footnotesize
\begin{tabular}{ccccc}
\hline\hline
C.o.g.&$791\pm124$&$574\pm40$&$878\pm73$&$307\pm180$\\
Name&$a_0(980)$&$K^*_0(700)$& $f_0(980)$ & $f_0(500)$\\
Mass&$980\pm20$ \cite{Zyla:2020zbs}\hspace{-1mm}&\hspace{-1mm}$648\pm 7$ \cite{Pelaez:2020gnd} \hspace{-1mm}&\hspace{-1mm}$996\pm7$ \cite{Pelaez:2019eqa}\hspace{-1mm}&\hspace{-1mm}$457\pm8$ \cite{Pelaez:2019eqa}\\ \hline\hline
RPP&$a_0(1450)$&$K^*_0(1430)$ & $f_0(1710)$ & $f_0(1370)$\\
\hline\hline
\end{tabular}
\renewcommand{\arraystretch}{1.}
\end{table}
\begin{table}[pb]
\vspace{-3mm}
\caption{\label{multiplet}Suggested spin-multiplet assignment for light mesons.\vspace{-2mm}}
\begin{center}
\renewcommand{\arraystretch}{1.3}
\begin{tabular}{cccc}
\hline\hline$a_2(1320)$ & $K^*_2(1430)$ & $f_2'(1525)$  & $f_2(1270)$\\
$a_1(1260)$ &  $K_{1A}$        & $f_1(1420)$  & $f_1(1285)$\\
$b_1(1235)$ & $K_{1B}$        & $h_1(1415)$ & $h_1(1170)$\\
$a_0(980)$  & $K^*_0(700)$ & $f_0(980)$ & $f_0(500)$\\
\hline\hline
\end{tabular}
\renewcommand{\arraystretch}{1.}
\end{center}
\vspace{-2mm}
\end{table}

In Table~\ref{assign} we compare the masses calculated using the c.o.g. rule
with the masses of the lowest-mass scalar mesons. The comparison is very surprising:
the predicted masses are rather close to the masses of the light scalar mesons.
The comparison suggests that the light scalar mesons are not only dynamically
generated. At the same time they might also play the role of ground-states of the scalar 
$q\bar q$ mesons. The Particle Data
Group interprets the mesons $a_0(1450)$, $f_0(1710)$, $f_0(1370)$, and $K^*_0(1430)$
as ground states of scalar mesons. This is clearly in conflict with the c.o.g. rule. 
Scalar mesons are subject to strong unitarization effects,
hence significant mass shifts can be expected. But in the PDG interpretation, 
these effects would need to be extremely large. The masses
calculated from the c.o.g. rule fall only a little below the actual light-scalar meson-masses. 
The masses of light mesons, given in the second line in Table~\ref{assign},
are certainly better in agreement with the values calculated using the c.o.g. rule
than with the assignment made by the Particle Data Group given in the last line
of the Table. In Table~\ref{multiplet} we show the suggested multiplet assignment.

There is one remark to be made: the $a_0(980)$ is spin-partner of
$a_2(1320)$, the $f_0(980)$ is not the spin-partner of $f_2(1270)$ but of
$f_2(1525)$.  In Jaffe's model, 
$a_0(980)$ and $f_0(980)$ are both $n\bar n s\bar s$. Here, $a_0(980)$
has a $n\bar n$ seed, with $n\bar n=(u\bar u+d\bar d)/\sqrt 2$, that acquires a large $s\bar s$ component while
$f_0(980)$ has a $s\bar s$ seed acquiring a large $n\bar n$ component. Due to their mass very close to
the $K\bar K$ threshold, both resonances develop a large $K\bar K$ molecular component
and can thus be generated dynamically. 

The assignment of, e.g., $f_2(1320)$, $f_1(1285)$, $f_0(500)$, and $h_1(1170)$, to one 
spectroscopic multiplet, to excitations with $L$=1, $S$=1
coupling to $J$=2,1,0 and $L$=1, $S$=0, $J$=1, is at variance with the Godfrey-Isgur model \cite{Godfrey:1985xj}: 
In this model, the lowest-mass 
isoscalar meson is predicted to have a mass of 1090\,MeV, certainly below the $f_2(1270)$ mass 
but far above the $f_0(500)$ mass. The Bonn model~\cite{Ricken:2000kf},
that is based on instanton-induced interactions instead of an effective one-gluon exchange, 
predicts 665\,MeV (model B), in better agreement with the c.o.g. rule.

\begin{figure}[pt]
\begin{center}
\hspace{-4mm}\begin{overpic}[scale=0.42,,tics=10]{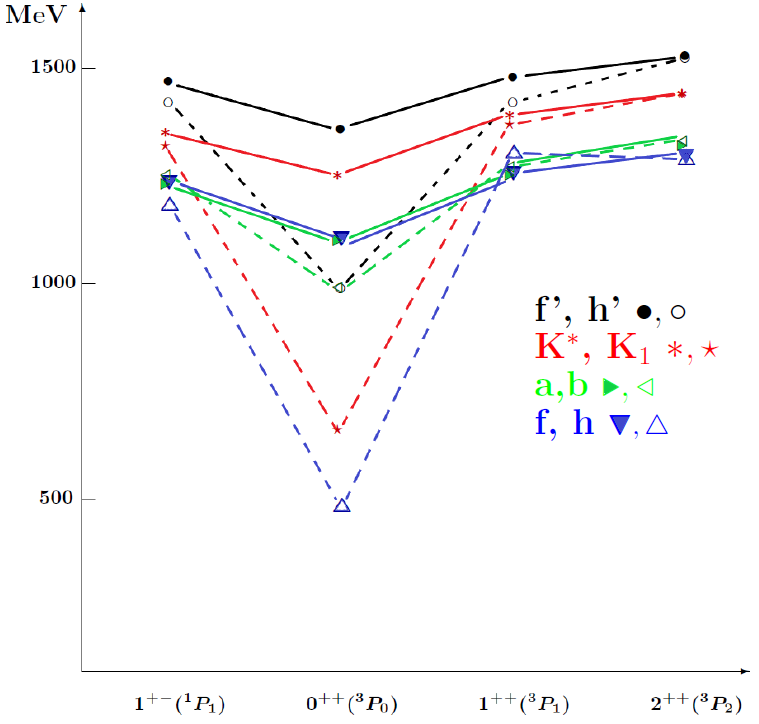}
\end{overpic}
\vspace{-4mm}
\end{center}
\caption{\label{firstexc}(Color online) The spectrum of $L=1$ excitations. Full symbols: Godfrey-Isgur model,
open symbols: RPP masses of suggested scalar spin-partners. The figure was suggested by W. Ochs.
}
\end{figure}

Figure~\ref{firstexc} shows a comparison of the spectrum of $L=1$ excitations as listed in
Table~\ref{multiplet} with the results of the Godfrey-Isgur model~\cite{Godfrey:1985xj}. 
The predicted masses of scalar mesons 
are all considerably above the scalar masses given in Table~\ref{multiplet}. However,
all predicted scalar masses fall below their multiplet partners. The RPP assignments to the lowest-mass
scalar $q\bar q$ mesons are all above their multiplet partners (except $K^*_0(1430)$ that is
degenerate in mass with $K^*_2(1430)$). Hence we believe that it is difficult to decide
on the basis of a quark model which scalar mesons belong to the $L=1$ multiplet.

Of course, the c.o.g. rule 
does not imply that the $f_0(500)$ is a pure $q\bar q$ state. It does not even need to have a
large $q\bar q$ component. A small $q\bar q$ seed could acquire
a strong $n\bar nn\bar n$ component. Color exchange 
makes this indistinguishable from a $\pi\pi$ component, the leading term when
the resonance is generated dynamically. Thus the light scalar mesons may have 
$q\bar q$, tetraquark and molecular contributions \cite{Close:2002zu}. 

We emphasize that the interpretation 
of the light scalar meson-nonet as part of the $q\bar q$ spectrum
is by no means in contradiction with chiral dynamics. This nonet is generated dynamically; but at the
same time it could play its part in the $q\bar q$ spectrum.    
This is one version of the well-known
chicken-and-egg problem: Is the interaction first, that generates the pole, or is the pole first, that generates
the interaction? QCD generates the interaction
{\it and} the pole, and one aspect is not thinkable without the other one.
The observation that the masses
of the light scalar mesons are rather close to the predictions based on the c.o.g. rule
opens the chance that these mesons could also be related the $q\bar q$ family. We are aware of the
fact that this interpretation is at variance with the modern understanding of the spectrum of
light scalar meson, but we have no other explanation why the c.o.g. rule ``happens" to be satisfied
for light mesons.

\section{\bf The flavor wave function}
\paragraph{\bf The mixing angle of pseudoscalar mesons} Mesonic mixing angles can be determined in 
the SU(3) singlet and octet basis or in a quark basis with $n\bar n$ and $s\bar s$.

For the pseudoscalar mesons, the difference in the octet and singlet decay constants 
is not negligible, and the nonet is described by two mixing angles 
$\theta_1^{\rm ps}$ and $\theta_8^{\rm ps}$~\cite{Feldmann:1998su}. Here, we neglect
the difference and use the singlet/octet basis in the form
\bc$
\left( \begin{array}{c}
\eta\\ 
\eta'\\
\end{array}\right)
=
\left( \begin{array}{cc}
\cos\theta^{\rm ps}& -\sin\theta^{\rm ps}  \\ 
\sin\theta^{\rm ps}  & \cos\theta^{\rm ps}\\
\end{array}\right)
\left( \begin{array}{c}
|8> \\ 
|1> \\
\end{array}\right)$
\qquad\qquad\ (6)\ec
and in the quark basis 
\bc$
\left( \begin{array}{c}
\eta\\ 
\eta'\\
\end{array}\right)
=
\left( \begin{array}{cc}
\cos\varphi^{\rm ps}& -\sin\varphi^{\rm ps}  \\ 
\sin\varphi^{\rm ps}  & \cos\varphi^{\rm ps}\\
\end{array}\right)
\left( \begin{array}{c}
|n\bar n> \\ 
|s\bar s> \\
\end{array}\right).$
\qquad\  \ (7)\ec
The ideal mixing angle $\theta_{\rm ideal}=35.3^\circ$ is defined by $\tan\theta_{\rm ideal}\\ =1/\sqrt2$
and leads to $\eta_{q\bar q}=-s\bar s$ and $\eta'_{q\bar q}=n\bar n$. In the quark
basis the two mixing angles  $\Phi_n^{\rm ps}$ and $\Phi_s^{\rm ps}$ are very similar in magnitude~\cite{Chen:2014yta},
we use  $\varphi^{\rm ps}=(39.3\pm1.0)^\circ$~\cite{Feldmann:1999uf}. In the singlet-octet basis, 
the unique mixing angle is given as 
$\theta^{\rm ps}=(39.3+35.3-90)^\circ= -(15.4\pm1.0)^\circ$, and
 the $\eta'$ is dominantly in the SU(3) singlet configuration. For $\theta^{\rm ps}=-15.5^\circ$,
\bc
$
\left( \begin{array}{c}
\eta  \\ 
\eta ' \\
\end{array}\right)
=
\left( \begin{array}{cc}
\sqrt{{13}/{14}} & +\sqrt{{1}/{14}}\ \    \\ 
-\sqrt{{1}/{14}}\ \   &\sqrt{{13}/{14}} 
\end{array}\right)
\left( \begin{array}{c}
|8> \\ 
|1> \\
\end{array}\right)$,\qquad\ 
(8)
\ec
and for $\varphi^{\rm ps}=39.2^\circ$ we get
\bc
$
\left( \begin{array}{c}
\eta \\ 
\eta'  \\
\end{array}\right)
=
\left( \begin{array}{cc}
\sqrt{{3}/{5}} & -\sqrt{{2}/{5}}   \\ 
\sqrt{{2}/{5}}   &\sqrt{{3}/{5}} 
\end{array}\right)
\left( \begin{array}{c}
|n\bar n> \\ 
|s\bar s> \\
\end{array}\right)$.
\qquad\quad\ (9)
\ec

\paragraph{\bf The mixing angle of scalar mesons} We now discuss the possibility to define the mixing angle for scalar mesons. 
For scalar mesons,
the refinement by two mixing angles exceeds the quality of data available at present, and
we use a mixing scenario with one mixing angle. In formulas we use $f_0(500)=\sigma$,
$K^*_0(700)=\kappa$, $a_0(980)=a_0$, and $f_0(980)=f_0$ as abbreviations.

We use the scalar mixing angles $\theta^{\rm s}$ and $\varphi^{\rm s}$ defined by
\bc$\hspace{-1mm}
\left( \begin{array}{c}
f_0\\ 
\sigma\\
\end{array}\right)
=
\left( \begin{array}{cc}
\cos\theta^{\rm s}& -\sin\theta^{\rm s}  \\ 
\sin\theta^{\rm s}  & \cos\theta^{\rm s}\\
\end{array}\right)
\left( \begin{array}{c}
|8> \\ 
|1> \\
\end{array}\right)$
\qquad\qquad\ \ \ (10)\ec
and
\bc$
\left( \begin{array}{c}
f_0\\ 
\sigma\\
\end{array}\right)
=
\left( \begin{array}{cc}
\cos\varphi^{\rm s}& -\sin\varphi^{\rm s}  \\ 
\sin\varphi^{\rm s}  & \cos\varphi^{\rm s}\\
\end{array}\right)
\left( \begin{array}{c}
|n\bar n> \\ 
|s\bar s> \\
\end{array}\right).$
\qquad\  \ \ \ (11)\ec
\setcounter{equation}{11}
This definition is used by Oller \cite{Oller:2003vf}, and $\varphi^{\rm s}=\theta^{\rm s}+(90-35.3)^\circ$.

The authors of Ref.~\cite{Anisovich:2001zp} use a wave function in the $q\bar q$ basis:
\beq
\hspace{-3mm}f_0=\sin\Phi^{\rm s} |s\bar s>+\cos\Phi^{\rm s} |n\bar n>
\eeq
with $\Phi^{\rm s}=-\varphi^{\rm s}$.

Ochs~\cite{Ochs:2013gi} assigns $f_0(980)$ and $f_0(1500)$ to the same multiplet and
determines the mixing angle. However, he uses only properties of $f_0(980)$ and
the decomposition
\beq
\hspace{-3mm}f_0=\cos\Phi'^{\rm s} |s\bar s>+\sin\Phi'^{\rm s} |n\bar n>
\eeq
with $\varphi^{\rm s}=\Phi'-90^\circ$. 
 
Most authors \cite{Li:2012sw,Aaij:2013zpt,Aaij:2014emv,Aaij:2015sqa,Liu:2019ymi,Soni:2020sgn}
use a definition of the mixing angle in the quark basis in the form 
\bc
$\hspace{-1mm}
\left( \begin{array}{c}
f_0  \\
\sigma  \\ 
\end{array}\right)
=
\left( \begin{array}{cc}
\cos\phi^{\rm s} & \sin\phi^{\rm s}  \\ 
-\sin\phi^{\rm s}  & \cos\phi^{\rm s}\\
\end{array}\right)
\left( \begin{array}{c}
|s\bar s> \\
|n\bar n> \\ 
\end{array}\right)$. \qquad\quad\ \  (14)\ec
\setcounter{equation}{14}
with with $\varphi^{\rm s}=\phi-90^\circ$.  

If the light scalar mesons behave like ordinary $q\bar q$ mesons, 
they should have a unique mixing angle.  Mixing angles can be derived from the production 
and decay of scalar mesons using SU(3) relations or from the 
masses exploiting the Gell-Mann--Okubo (GMO) formula.
Most authors neglect the effects of the different densities at the origin 
of the $f_0(500)$ and $f_0(980)$ wave functions and
assumed that these resonances are $q\bar q$ states: these are certainly two 
highly questionable assumptions. Nevertheless we will compare the mixing angles
derived using different methods.   

\paragraph{\bf The mixing angle from decays of scalar mesons}
Oller~\cite{Oller:2003vf} determined the mixing angle in the singlet-octet basis
of the light scalar mesons in a SU(3) analysis of the meson decay couplings. The coupling
constants were obtained by determining the residues at the pole positions of the resonances.
The analysis gave $\theta^{\rm s}=(19\pm5)^\circ$ or  $\varphi^{\rm s}= (74\pm5)^\circ$. 
Many determinations of the mixing angle
are ambiguous. In these cases, we choose the value that is closer to this value. The
resulting mixing angles are collected in Table~\ref{prod}.

The authors of Ref.~\cite{Anisovich:2001zp} study radiative decays of $\Phi$-mesons and
the two-photon width of scalar and tensor mesons. The calculated two-photon width of scalar
and tensor mesons agree well with data even though approximately equal radial wave
functions are used. For this reason, the authors argue that
$a_2(1320), f_2(1270)$ and $f_2(1525)$ might belong to the same $P$-wave multiplet
as $a_0(980)$ and $f_0(980)$. For the scalar mixing angle, two solutions were
found $\Phi^{\rm s}=-(48\pm6)^\circ$ or $(86\pm3)^\circ$. Equations (11) and (14)
are related by  $\Phi^{\rm s}=-\varphi^{\rm s}$. The latter angle
defines a mixing angle $\varphi^{\rm s}= -(86\pm3)^\circ$. The overall sign of the
wave function is not important here, hence we quote $\varphi^{\rm s}= (94\pm3)^\circ$ 
in Table~\ref{prod}.  

Ochs~\cite{Ochs:2013gi} determined $\phi^{\rm s}$ from the couplings of $f_0(980)$
to $\pi\pi$ and $K\bar K$, the two-photon widths of $f_0(980)$ and $a_0(980)$, from
the production rates for $D^+_{(s)}\to f_0\pi$ and $D^+\to K^*_0(1430)\pi^+$, and
from  $J/\psi\to f_0(980)\omega (\Phi)$. He found two solutions $\phi^{\rm s}= (30\pm3)^\circ$
and  $\phi^{\rm s}= (162\pm3)^\circ$.  Only the latter one is compatible with~\cite{Oller:2003vf}
and corresponds to $\varphi=(72\pm3)^\circ$.

In Ref.~\cite{Hanhart:2007wa}, the use of the 
density of the wave function at the origin for hadronic molecules is critisized; instead, dressed meson propagators 
and photon emission vertices are used to calculate the two-photon widths of $f_0(980)$.
The result is in excellent agreement with the experimental value. The 
radiative $\Phi$ decays into $f_0(980)$ and $a_0(980)$ are also consistent with the molecular
view of these mesons~\cite{Kalashnikova:2004ta}.

\paragraph{\bf The mixing angle from the production of scalar mesons}

Li, Du, and L\"u~\cite{Li:2012sw} deduce the mixing angle from the rate $(1.2\pm0.3)\cdot10^{-4}$
for $B_s^0 \to J/\psi f_0(980)$ decays and find $\phi^{\rm s} =(34^{+\ 9}_{-15})^\circ $ 
or $\phi^{\rm s} =(146^{+15}_{-\ 9})^\circ$. The latter value yields $\varphi=(56^{+15}_{-\ 9})^\circ$.

In a study of  
the reaction $B^0 \to \overline{D}^0 \pi^+\pi^-$~\cite{Aaij:2015sqa}, the collaboration
found significant contributions from $f_0(500)$ and $f_0(980)$. In this process, scalar mesons
are produced (and decay) due to their $n\bar n$ component. The production ratio is
interpreted within a $q\bar q$ and tetraquark mixing scheme. Assuming identical form factors 
for both mesons, the $q\bar q$ model gives $\tan^2\phi^{\rm s}=0.177^{+0.066}_{-0.062}$ 
with $\phi^{\rm s}=\pm(23\pm4)^\circ$. We choose the negative sign and get 
$\varphi^{\rm s}=-(113\pm4)^\circ$ or, changing the overall sign, $\varphi^{\rm s}=(67\pm4)^\circ$.

The LHCb collaboration also studied the reaction $\bar{B}_s^0\to J/\psi\pi^+\pi^-$ 
\cite{Aaij:2014emv}. The $\pi^+\pi^-$ invariant mass spectrum shows a large contribution
from $f_0(980)$ and no sign for $f_0(500)$. From the upper limit, the scalar mixing angle 
is constrained to $|\phi^{\rm s}|<7.7^\circ$ at 90\% confidence level, or $(3\pm3)^\circ$. 
We use $\varphi^{\rm s}=-(87\pm3)^\circ$.

\begin{table}[pt]
\caption{\label{prod}Scalar mixing angle $\varphi^{\rm s}$ from production experiments and
from $\sigma$ decay coupling constants \cite{Oller:2003vf}. The sign
of the mixing angles is mostly undetermined. It is chosen to be compatible with Ref.~\cite{Oller:2003vf}.
}
\begin{center}
\small
\renewcommand{\arraystretch}{1.5}
\begin{tabular}{ccccccc}
\hline\hline
 $(74\pm5)^\circ$\hspace{-3mm}&\hspace{-2mm}(94\er6)$^\circ$&\hspace{-3mm}$(72\pm3)^\circ$\hspace{-3mm}&\hspace{-3mm}$(56^{+15}_{-\ 9})^\circ$\hspace{-2mm}&\hspace{-2mm}(67\er4)$^\circ$\hspace{-1mm}&\hspace{-1mm}(87\er3)$^\circ$\\
\hspace{-3mm}\cite{Oller:2003vf}&\hspace{-3mm}\cite{Anisovich:2001zp}&\hspace{-6mm}\cite{Ochs:2013gi}\hspace{-3mm}&\hspace{-3mm}\cite{Li:2012sw}&\hspace{-3mm}\cite{Aaij:2015sqa}&\hspace{-3mm}\cite{Aaij:2014emv} \\
\hline\hline
\end{tabular}
\begin{tabular}{ccccc}
$\approx$65$^\circ$&(58\er8)$^\circ$\hspace{-2mm}&\hspace{-2mm}$(67\pm3)^\circ$\hspace{-3mm}&\hspace{-5mm}$(55\pm5)^\circ$\hspace{-3mm}&\hspace{-3mm}$(63\pm2)^\circ$\\
\cite{Liu:2019ymi}&\hspace{-3mm}\cite{Soni:2020sgn}&\hspace{-3mm}$D^+$$\to$$\pi^+\pi^+\pi^-$\hspace{-3mm}&\hspace{-3mm}$D^0$$\to$$\pi^0\pi^+\pi^-$\hspace{-3.5mm}&\hspace{-3mm}$J/\psi$$\to$$\gamma \pi\pi$\\\hline\hline
\end{tabular}
\renewcommand{\arraystretch}{1.}
\end{center}
\vspace{-2mm}
\end{table}

The authors of Ref.~\cite{Liu:2019ymi} study $B_{d,s}^0\to J/\psi f_0(500)$ $[f_0(980)]$ decays.
In these decays, the~$\bar b$-quark in the $B_{d,s}$ converts into a $\bar c$ quark under emission of
a $W$-boson. The $W$ decays into a $c$ quark plus a $\bar d$ -- picking up the $d$-quark of the
$B_{d}$ -- or an $\bar s$ that combines with the $s$-quark of the $B_{s}$. The squared transition
amplitudes are thus proportional to the $n\bar n$ or $s\bar s$ content of the scalar wave
functions. Assuming similar hadronizations from the primary $q\bar q$ pair into 
the two scalar mesons, a scalar mixing angle in the $n\bar n$-$s\bar s$ basis 
of $|\phi^{\rm s}|\approx 25^\circ$ or $\varphi^{\rm s}=65^\circ$ is deduced.  

The authors of Ref.~\cite{Soni:2020sgn} compared
the leptonic decays of the charmed mesons $D^+$ into $f_0(980)/a_0(980)+e^+\nu_e$, 
$D^0$ into $a_0(980)+e^+\nu_e$, and $D^+_s$ into $f_0(980)+e^+\nu_e$. 
The scalar mixing angle was $|\phi^{\rm s}|$ shown to be
compatible with values in the  (25 - 40)$^\circ$ range, a result that we use in the
form  $\varphi^{\rm s}=(58\pm8)^\circ$.

We add a few further determinations of the scalar mixing angle. The errors are of statistical nature only. 
Dalitz-plot analyses to $D^+$~\cite{Aitala:2000xu,Bonvicini:2007tc} and $D^0$~\cite{Aubert:2007ii} decays
into three pions reveal the fractional contributions of $f_0(500)$ and $f_0(980)$. The mean values
for the $D^+$ contributions from Refs.~\cite{Aitala:2000xu,Bonvicini:2007tc} are 
(0.422\er0.027)\% for $f_0(500)$ and (0.048\er\ 0.010)\% for $f_0(980)$. 
For $D^0$ decays the contributions (0.82\er0.10 \er0.10)\% and (0.25\er0.04\er0.04)\% are 
reported \cite{Aubert:2007ii}. The $f_0(500)$
contribution is proportional to $\cos^2\phi^{\rm s}$, the $f_0(980)$ contribution to $\sin^2\phi^{\rm s}$. 
Using the unweighted mean and spread
of RPP results on $\Gamma_{f_0(980)\to \pi\pi}/\Gamma_{\rm tot}=0.72\pm0.03$ and after
correction for the different phase spaces, mixing angles are determined that are listed in Table~\ref{prod}. 

From the ratio of $f_0(500)$ and $f_0(980)$ production in the reaction
$\overline{B}^0 \to J/\psi\pi^+\pi^-$, the LHCb collaboration constrained
the mixing angle $|\phi^{\rm s}|$ to be $<31^\circ$~\cite{Aaij:2013zpt} at 90\%
confidence level.  With increased statistics~\cite{Aaij:2014siy}, the LHCb collaboration
confirmed the absence of $f_0(980)$ production in this reaction. A reanalysis of the data
exploiting a dispersive framework demonstrated a significant role of $f_0(980)$ in the 
$\overline{B}^0 \to J/\psi\pi^+\pi^-$ reaction~\cite{Daub:2015xja}. Therefore we exclude
this reaction in the present discussion. 
The ratio of the two frequencies $f$ for $J/\psi\to\omega f_0(980)$, 
$f$=(1.4\er0.5)\,$10^{-4}$, and  $J/\psi\to\phi f_0(980)$, 
$f$=(3.2\er0.9)\,$10^{-4}$ \cite{Zyla:2020zbs} gives a compatible
mixing angle that is not used since this ratio is included in the evaluation by Ochs~\cite{Ochs:2013gi}.

The  frequencies for $J/\psi\to\gamma f_0(500)$, with $f$=(11.4\er2.1) $10^{-4}$, and 
$J/\psi\to \gamma f_0(980)$, with $f$=(0.21\er 0.04)\,$10^{-4}$ \cite{Sarantsev:2021ein}, 
can be used to determine the mixing angle $\theta^{\rm s}$=$\pm$(8.0$\pm$1.2)$^\circ$
that leads to $\phi^{\rm s}=-(27.3\pm1.2)^\circ$ or $-(43.3\pm1.2)^\circ$. The former
value translates into $\varphi^{\rm s}=-(62.7\pm1.2)^\circ$ that is compatible the value obtained in 
Ref. \cite{Oller:2003vf}.
Here it is assumed, that the radiative $J/\psi$ decay couples only to the SU(3) singlet components.

The eleven mixing angles are statistically not compatible. We assume the uncertainties
are dominated by systematic uncertainties. Therefore, we take the mean value of the
scalar mixing angles. The full spread is $11^\circ$ that corresponds to a statistical uncertainty of  
$4^\circ$. The mixing angle $\theta^{\rm s}$ in the singlet/octet basis and the quark basis 
$\phi^{\rm s}$ are related by $\phi^{\rm s}=\theta^{\rm s}+90-35^\circ$.  
Thus we have
\be
\varphi^{\rm s}=(69\pm4)^\circ \qquad \theta^{\rm s}=(14\pm4)^\circ.
\ee 
For $\theta^{\rm s}=14.4^\circ$, the octet contribution to the $f_0(500)$
and the singlet contribution to the $f_0(980)$ are small: 
\bc
$
\left( \begin{array}{c}
f_0  \\
\sigma  \\ 
\end{array}\right)
=
\cdot \left( \begin{array}{cc}
\ \ \sqrt{\frac{15}{16}} & +\sqrt{\frac{1}{16}} \\ 
-\sqrt{\frac{1}{16}}  &\ \ \sqrt{\frac{15}{16}}\\
 \end{array}\right)
\left( \begin{array}{c}
|8> \\
|1> \\ 
\end{array}\right)$.
\ec
In the quark basis and for $\phi^{\rm s}=69.3^\circ$, the wave functions can be cast into the  form
\bc
$
\left( \begin{array}{r}
f_0  \\
\sigma  \\ 
\end{array}\right)
=
\left( \begin{array}{rr}
\sqrt{{1}/{8}} & -\sqrt{{7}/{8}}   \\ 
\sqrt{{7}/{8}}   &\sqrt{{1}/{8}} 
\end{array}\right)
\left( \begin{array}{r}
|n\bar n> \\ 
|s\bar s> \\
\end{array}\right)$,
\ec
and  $f_0(500)$ is mostly an $n\bar n$ state and $f_0(980)$ 
an $s\bar s$ state.
 
\paragraph{\bf The flavor wave function from the GMO formula}
The mass pattern of the lightest scalar mesons differs decisively from the one
observed for other mesons. There is an isovector meson $a_0(980)$ and an
isoscalar meson $f_0(980)$ -- like $\rho$ and $\omega$ -- but in contrast to the
vector mesons, these scalar mesons are heavier than their nonet partners. 
The other isoscalar meson, $f_0(500)$, has the lightest mass of the multiplet.
The spectrum of scalar mesons resemble the pseudoscalar mesons but inverted:
There are the low-mass $\pi$ and $\eta$ that correspond to $a_0(980)$ and $f_0(980)$,
the $K$ and $K^*_0(700)$, and the high-mass $\eta'$ corresponds 
to the low-mass $f_0(500)$.

The sum rule 
\beq
(m_f + m_{f'})(4 m_K -m_a)-3m_fm_{f'} &(0.977\,{\rm GeV}^2)\nonumber\\
=8m_k^2-8m_Km_a+3m_a^2 &(1.16\,{\rm GeV}^2)\nonumber
\eeq
contains $f$ and $f'$ symmetrically. It is violated at the 6\% level when the 
difference is compared to $(8m_k^2+3m_a^2)$. 
We identify $\sigma$ as $f$ and $f_0$ as $f'$. 

Also the  equal splitting law
\beq
m_\kappa ^2 - m_K^2 =0.172= m_\sigma ^2 - m_\pi^2= 0.191\ {\rm GeV}^2
\eeq
suggested by Scadron~\cite{Scadron:1992sp} is satisfied at the 10\% level.

For ideal mixing, we expect 
\beq
m_\kappa =\frac{m_{f_0} + m_\sigma}{2}.
\eeq
The expectation holds true at this 10\% level. This is in line with the linear GMO formula
\beq
\tan\theta = \frac{4m_K-m_a-3m_{f'}}{2\sqrt{2}(m_a-m_K)}=-1.465
\label{GMO-lin}
\eeq
that yields a negative mixing angle, $\theta_{\rm lin}^{\rm s}=-(55.7\pm2.5)^\circ$,
$\phi_{\rm lin}^{\rm s}=-(91.0\pm2.5)^\circ$.

Alternatively we use the mass formula in the form
\beq
\tan^2\theta =\frac{4m_K-m_a-3m_{f'}}{-4m_K+m_a+3m_f}=5.782
\eeq
and obtain $\theta_{\rm lin}^{\rm s}=\pm(67.4\pm 2.1)^\circ$. For the negative sign we obtain
$\phi=-(102.7\pm2.1)^\circ$. The value $\phi=-90^\circ$ would indicate an inverted spectrum.
Replacing the masses by squared masses in eqn.~(\ref{GMO-lin}) yields  
$\theta_{\rm quad}^{\rm s}=-(55.9\pm2.5)^\circ$ and $\phi_{\rm quad}^{\rm s}=-(91.2\pm2.5)^\circ$. 
In Table~\ref{GMO} we  compare the mixing angles of scalar and pseudoscalar mesons.

The GMO formula suggests that $f_0(500)$ is mostly an $n\bar n$ state and $f_0(980)$ 
an $s\bar s$ state.  From the production and decay of light scalar mesons we conclude that
$f_0(500)$ is mainly in the singlet, $f_0(980)$ mainly in the octet configuration, 
with a small mixing angle. 

There is an important difference between the mixing angle derived from production
and decay of mesons or from the GMO formula. Production and decay depend
on the mesonic wave function, the GMO formula if sensitive to the mass content. 

\begin{table}[pt]
\caption{\label{GMO}Pseudoscalar and scalar mixing angles from the GMO formula.
The pseudoscalar mixing angles are from Ref.~\cite{Zyla:2020zbs}.}
\begin{center}
\renewcommand{\arraystretch}{1.3}
\begin{tabular}{lllllll}\hline\hline
$\theta_{\rm lin}^{\rm\,ps}$&\hspace{-3mm}=\hspace{-3mm}&$-24.5^\circ$ &\qquad& $\phi_{\rm lin}^{\rm s}$&\hspace{-3mm}=\hspace{-3mm}&$+(91.0\pm 2.5)^\circ$\\
$\theta_{\rm quad}^{\rm\,ps}$&\hspace{-3mm}=\hspace{-3mm}&$-11.3^\circ$ && $\phi_{\rm quad}^{\rm s}$&\hspace{-3mm}=\hspace{-3mm}&$+(91.2\pm2.5)^\circ$\\
\hline\hline
\end{tabular}
\renewcommand{\arraystretch}{1.}
\end{center}
\vspace{-3mm}
\end{table}

With the caveats expressed above ($f_0(500)$ and $f_0(980)$ may have different wave functions
and are likely no $q\bar q$ states) we can still note that a mixing angle for the light scalar
mesons can be defined. The $f_0(500)$ is a mainly $n\bar n$ state but has an additional
singlet component, beyond the one expected from the decomposition of $n\bar n$ into 
singlet and octet. The $f_0(980)$ mainly $s\bar s$ with an additional octet component.

\paragraph{\bf Size of light scalar mesons}
Weinberg derived a criterium to decide if a bound state is elementary or compact
or if it is an extended particle~\cite{Weinberg:1962hj,Matuschek:2020gqe}. A quantity $Z$ can be defined
that is related to the scattering length $a$ via
\be
a &=& -2\,\frac{1-Z}{2-Z}\frac{1}{\gamma}
\ee
where $\gamma =\sqrt{2\mu E_B}$ denotes the binding momentum. The quantity $Z$
determines the compact fraction of the bound state, $1-Z$ the ``molecular" fraction.
For $Z=0$, the system is ``molecular", for $Z=1$ compact. 

The authors of Ref.~\cite{Guo:2015daa} extended this relation and derived 
a probabilistic interpretation of the compositeness relation for resonances.
Taking the effective range into account, the authors 
determined $Z_{\sigma}$  to $0.60\pm0.02$ where the error is of
statistical nature only. The smallness of the molecular fraction in the $f_0(500)$
wave function is in line with a determination of the rms radius of the $f_0(500)$
from a Taylor expansion of the $f_0(500)$ form factor~\cite{Albaladejo:2012te}: 
\beq
r_{\rm rms}^\sigma= [(0.44\pm0.03)-i(0.07\pm0.03)]\,{\rm fm}.
\eeq
The $f_0(500)$ meson seems to be a compact object.  

A similar analysis was reported in Ref.~\cite{Baru:2003qq} for $a_0(980)$ and $f_0(980)$. The authors concluded 
that the probability to find the $a_0(980)$ as $q\bar q$ state is about 25 to 50\% and
for the $f_0(980)$ meson this probability is even smaller, about 20\% or less. 
In Ref.~\cite{Guo:2015daa}, the probability to find the $f_0(980)$ as $q\bar q$ state 
is $Z_{f_0}=0.33^{+0.28}_{-0.28}$. The compact fractions of $f_0(500)$ and $f_0(980)$ differ 
significantly, even though the large errors do not completely exclude similar spatial wave 
functions of $f_0(500)$ and $f_0(980)$. The $K^*_0(700)$ has only a small molecular 
component, $Z_{\kappa}=0.88$.

Obviously, the Weinberg criterium does not exclude a possibly small $q\bar q$ component in the
wave functions of light scalar mesons.

\section{Excited scalar states}
\begin{table}[pb]
\caption{\label{radex}Scalar mesons: ground states and radial excitations. Masses and the assignment
to mainly-singlet and octet configurations stems from Ref.~\cite{Sarantsev:2021ein}.  }
\begin{center}
\renewcommand{\arraystretch}{1.3}
\begin{tabular}{cccc}
\hline\hline
$I=1$ & $=1/2$& $I=0$& $I=0$\\[-1ex]
         &            & ``octet"& ``singlet"\\\hline
 $a_0(980)$ & $K^*_0(700)$  & $f_0(980)$ & $f_0(500)$ \\
$a_0(1450)$ & $K^*_0(1430)$& $f_0(1500)$ & $f_0(1370)$\\
                 &                        & $f_0(1770)$ & $f_0(1710)$\\
$a_0(2020)$  & $K^*_0(1950)$& $f_0(2100)$&$f_0(2020)$\\
                 &                        & $f_0(2330)$ & $f_0(2200)$\\
\hline\hline
\end{tabular}
\renewcommand{\arraystretch}{1.}
\end{center}
\end{table}

When $a_0(1450)$, $K^*_0(1430)$, $f_0(1370)$, and $f_0(1710)$ are not elements of the
lowest-mass $q\bar q$ multiplet with $L$\,=\,1, do they fit into a radial excitation multiplet? 
Unfortunately, only few states are known to complete the $q\bar q$ multiplets with $L=1$ and
one unit of radial excitation. With the masses of $a_2(1700)$, $a_1(1640)$, and $a_0(1450)$
as given in the RPP, we predict -- using the c.o.g. rule -- a $b_1$ mass of $M_{b_1(xxx)}= (1663\pm 23)$\,MeV.
Quark models predict higher masses for the radial excitations, hence we use mass differences relative  
to the $a_2(1700)$ mass. The Godfrey-Isgur~\cite{Godfrey:1985xj} (Bonn~\cite{Ricken:2000kf}) 
model predicts the $b_1$ mass to be 40 (50)\,MeV below the $a_2(1700)$ mass; both values are
in excellent agreement with the c.o.g. prediction.    
The $f_2(1640)$, $h_1(1595)$, $f_0(1370)$ can be used to predict the mass of a $f_1(xxx)$ to 
$1567^{+60}_{-92}$\,MeV. The Godfrey-Isgur~\cite{Godfrey:1985xj} (Bonn~\cite{Ricken:2000kf})
model predicts the $f_1$ mass to be 40 ( $\approx60$)\,MeV below the $f_2$ mass, certainly compatible
with the predicted mass. According to the c.o.g. rule, the $a_0(1450)$ resonance is a credible candidate
to be the first radial excitation of the $a_0(980)$ and to be spin-partner of $a_2(1700)$, $a_1(1640)$ and
an unobserved $b_1$ expected at 1655\,MeV.  The
 $f_0(1370)$ could be the radial excitation of $f_0(500)$ and spin-partner of
$f_2(1640)$, $h_1(1595)$, and a missing $h_1$ at about 1590\,MeV.
There is neither a $a_0$ nor 
a $K^*_0$ state that could be partner of $f_0(1770)$/$f_0(1710)$
or of $f_0(2330)$/$f_0(2200)$.

In Table~\ref{radex} we collect the known scalar mesons. Three full nonets can 
be defined. The GMO formula suggest a mixing angle for the four mesons
$a_0(1450)$, $K^*_0(1430)$, $f_0(1500)$, $f_0(1370)$ that is nearly ideal. The mixing angle for 
$a_0(2020)$, $K^*_0(1950)$, $f_0(2100)$, and $f_0(2020)$ vanishes in the singlet/octet basis. 
In both cases the errors are too large to make this a solid statement.
\section{The fragmented glueball}

QCD predicts the existence of glueballs, of states without constituent quarks. Here we quote a few calculations of the glueball 
spectrum~\cite{Bali:1993fb,Morningstar:1999rf,Szczepaniak:2003mr,Gregory:2012hu,Athenodorou:2020ani,Huber:2020ngt,Rinaldi:2021dxh}. 
The lowest-mass glueball is expected to
have scalar quantum numbers, a mass in the 1500 to 2000\,MeV range, to intrude the spectrum
of scalar mesons and to mix with them. A large number of different mixing schemes were reported
initiated by the work of Amsler and Close~\cite{Amsler:1995tu,Amsler:1995td}. The proof for
the presence of a glueball is supposed to be supernumerosity: it is expected that more mesons
should be observed than predicted by quark models. 

\begin{figure}[pt]
\begin{center}
\begin{overpic}[scale=0.36,,tics=10]{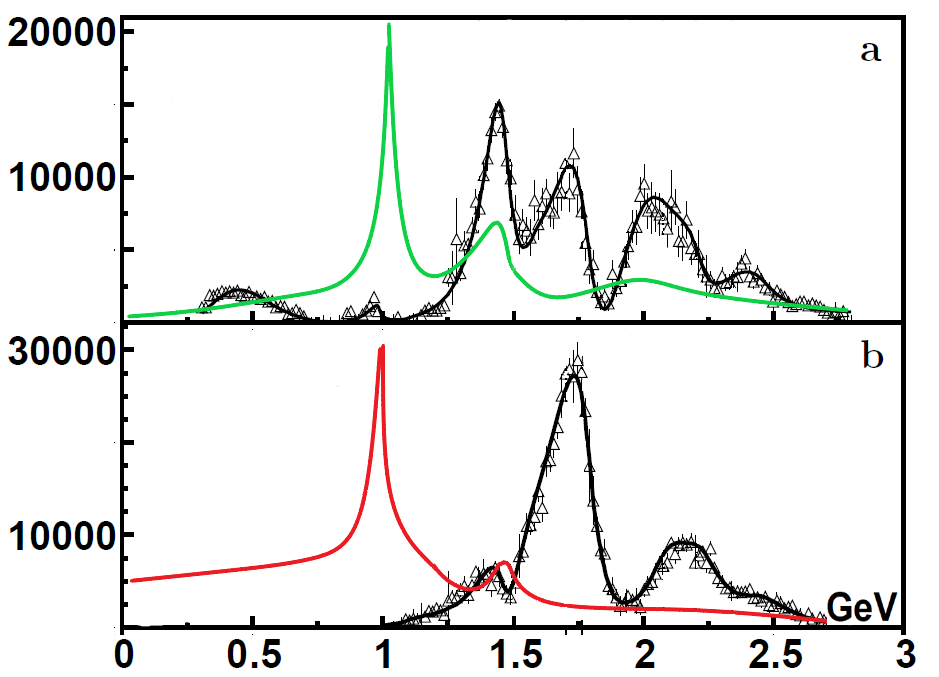}
\put(16,64){\large\boldmath $gg\to\pi\pi$ }
\put(16,59){\large\boldmath{\gr $s\bar s\to\pi\pi$}}
\put(16,31){\large\boldmath $gg\to K\bar K$ }
\put(16,26){\large\boldmath{\rd $s\bar s\to K\bar K$}}
\end{overpic}\vspace{-2mm}
\end{center}
\caption{\label{Data}(Color online) $\pi^0\pi^0$ (a)~\cite{Ablikim:2015umt} and $K_SK_S$ (b)~\cite{Ablikim:2018izx} invariant mass
distributions from radiative $J/\psi$ decays (histogramm) with fit~\cite{Sarantsev:2021ein}  
and the pion (a) and kaon (b) formfactors~\cite{Ropertz:2018stk} from $\bar B_s^0\to J/\psi\pi^+\pi^-$~\cite{Aaij:2014emv} and
$\bar B_s^0\to J/\psi K^+K^-$~\cite{Aaij:2017zgz}. \vspace{6mm}
}
%\end{figure}
%\begin{figure}
\begin{tabular}{cc}
\hspace{6mm}\begin{overpic}[scale=0.26]{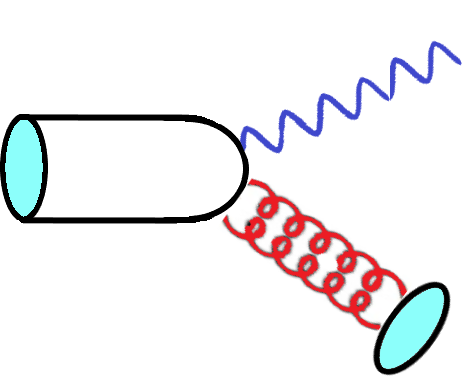}
\put(20,59){\large\boldmath $\bar c$ }
\put(20,25){\large\boldmath $c$ }
\put(-30,40){\large\boldmath $J/\psi$ }
\put(100,2){\large\boldmath{\rd $f_0$}}
\put(100,70){\large\boldmath{\bl $\gamma$}}
\end{overpic}&
\hspace{6mm}\begin{overpic}[scale=0.26]{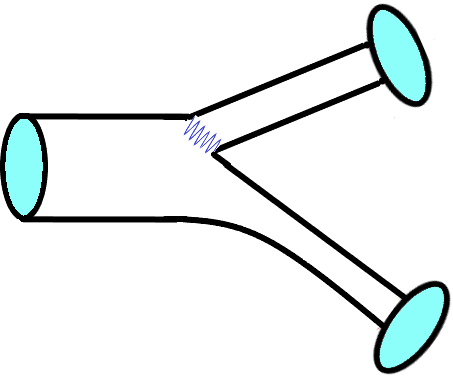}
\put(20,60){\large\boldmath $\bar b$ }
\put(40,25){\large\boldmath $s$ }
\put(65,50){\large\boldmath $c$ }
\put(65,71){\large\boldmath $\bar c$ }
\put(65,37){\large\boldmath $\bar s$ }
\put(25,47){\large\boldmath\footnotesize $W^+$ }
\put(-21,40){\large\boldmath $\bar B^0_s$ }
\put(100,2){\large\boldmath{\rd $f_0$}}
\put(100,70){\large\boldmath{\bl $J/\psi$}}
\end{overpic}\vspace{2mm}
\end{tabular}
\caption{\label{Reactions}(Color online) In radiative $J/\psi$ decays, two gluons,  in $\bar B^0_s\to J/\psi +s\bar s$, a
$s\bar s$ pair may convert into a scalar meson. \vspace{-30mm}
}
\end{figure}

This expectation was not met in a recent fit to a large body of reactions on radiative $J/\psi$ decays,
$\pi\pi$ elastic scattering, pion-induced reactions and $\bar pp$ annihilation~\cite{Sarantsev:2021ein}.
The multi-channel fit to the data revealed the existence of ten scalar isoscalar resonances from 
$f_0(500)$ to $f_0(2330)$. Of particular importance for the interpretation are the data on
radiative $J/\psi$ decays to $\pi^0\pi^0$~\cite{Ablikim:2015umt} and $K_SK_S$~\cite{Ablikim:2018izx}
shown in Fig.~\ref{Data}. 
In these reactions, the $J/\psi$ is supposed to decay into one photon and two gluons (see Fig.~\ref{Reactions}, left). The two
gluons interact and undergo hadronization. 

Figure~\ref{Data} compares the invariant mass distributions resulting from radiative $J/\psi$ decays with the
pion and kaon form factors. Their square is proportional to the cross sections. The form factors were deduced by 
Ropertz, Hanhart and Kubis~\cite{Ropertz:2018stk} 
exploiting the reactions $\bar B_s^0\to J/\psi$ $\pi^+\pi^-$~\cite{Aaij:2014emv} and
$\bar B_s^0\to J/\psi K^+K^-$~\cite{Aaij:2017zgz} and imposing the $S$-wave phase shifts from $\pi\pi$ and $K\bar K$.
 In the $\bar B^0_s$ decays, a primary $s\bar s$ pair converts
into the final state mesons  (see Fig.~\ref{Reactions}, right). The scale of the formfactor is chosen to match the
intensity at high masses. 

Both formfactors are dominated by the $f_0(980)$ resonance. The resonance connects the initial $s\bar s$ pair
to $\pi^+\pi^-$ in the final state: the $f_0(980)$ must have $s\bar s$ and $n\bar n$ components, and this statement 
holds true for the $f_0(1500)$ as well. The $f_0(980)$ is nearly absent in radiative $J/\psi$ decays: this was explained
by a dominance of the SU(3) octet component  in  $f_0(980)$.  Most striking is the mountain landscape
above 1500\,MeV in the data on radiative $J/\psi$ decays. The huge peak, e.g., at 1750\,MeV in the $K\bar K$ 
mass spectrum and the smaller one at 2100\,MeV decay prominently into $K\bar K$ but are produced with  
two gluons in the initial state and not -- at least not significantly -- with $s\bar s$ in the initial state. This is highly remarkable: the two
gluons in the initial state must be responsible for the production

\qquad\phantom{z}\vspace{25mm}\\\noindent
of resonances that decay strongly into $K\bar K$
but are nearly absent when $s\bar s$ pairs are in the initial state. Also the rich structure in the radiatively
produced $\pi\pi$ mass spectrum is accompanied with little activity when the initial state is $s\bar s$. The 
rich structure stems from gluon-gluon dynamics. 

In Ref.~\cite{Sarantsev:2021ein} it was shown that the total
yield of scalar mesons as a function of their mass can be described by a Breit-Wigner function with mass
$M_G=(1865\pm 25^{\,+10}_{\,-30})\,{\rm MeV}$ and width $\Gamma_G= (370\pm 
50^{\,+30}_{\,-20})\,{\rm MeV}$. The peak is created by gluon-gluon interactions. It is the scalar glueball.
The glueball peak extends over several scalar resonances, the glueball is fragmented. These scalar mesons 
can be grouped into a class of mesons with  
{\it mainly octet} and {\it mainly singlet} $q\bar q$ components. The two classes of mesons fall onto two linear $(n, M^2)$ 
trajectories, there is no supernumerous state.

The classification into mainly singlet and mainly octet resonances is based on the opposite interference 
pattern in $\pi\pi$ and $K\bar K$ (see Ref.~\cite{Sarantsev:2021ein}). This observation seems to
suggest that the scalar resonance in in the mass range from 1700 to 2100\,MeV are produced largely
due to their $gg$ component in the wave function while they decay largely via their $q\bar q$ component.

\section{Summary and conclusions}

In this paper, an alternative view of scalar mesons was presented. The light scalar mesons
are interpreted as members of a spin multiplet that includes the well known tensor and axial vector
mesons. The scalar glueball of lowest mass is not seen as supernumerous state invading the spectrum
of scalar isoscalar states and mixing with them but rather as component of the wave functions of scalar
isoscalar mesons.  

These two conjectures are linked. The wave function of a scalar meson can be expanded
\be
\hspace{-6mm}{\rm Meson\ } &=& \alpha_1 |q\bar q>+ \alpha_2 |qq\bar q\bar q> + \alpha_3 |gg>   \\&&+ 
\alpha_4 |{\rm meson\hspace{-1mm}-\hspace{-1mm}meson}> + \alpha_5 |q\bar qg> + \cdots.\nonumber
\ee
with possible contributions from higher terms. We thus could expect five (or more) different types of states. 
Since they all have the same quantum numbers, they can mix. But the number of states to be expected
is large. We now make a conjecture: we assume that only one state exist and that the orthogonal states
disappear in the continuum. Low-mass scalar mesons can be dynamically generated but still have a $q\bar q$ seed.
Scalar mesons can have a large glueball component but still are part of the regular spectrum of scalar $q\bar q$
mesons.    

The situation can be compared to
the one observed in case of the $N(1535)1/2^-$ resonance. This resonance
is generated dynamically~\cite{Kaiser:1995cy}
 but still plays an important role in quark models as member of the nucleon's first
excitation multiplet with orbital angular momentum $L=1$. It may also be interpreted as 
pentaquark state with hidden strangeness~\cite{Zou:2007mk}.
These different views, to consider $N(1535)1/2^-$ as
dynamically generated from meson-baryon interactions,
its interpretation as pentaquark, {\it and} its assignment to the 
three-quark baryon spectrum  are legitimate and provide additional insights. 
The light scalar mesons can be dynamically generated molecules but they could still
have a $q\bar q$ seed and play their part in quark models.\\[2ex]

{\it 
I would like to thank Chr. Hanhart, W. Ochs, J. Oller, and U. Thoma for a careful reading of the manuscript and 
helpful discussions. Funded by the NSFC and the Deutsche Forschungsgemeinschaft (DFG, German 
Research Foundation) through the funds provided to the Sino-German Collaborative
Research Center TRR110 “Symmetries and the Emergence of Structure in QCD”
(NSFC Grant No. 12070131001, DFG Project-ID 196253076 - TRR 110) }

\end{document}